\documentclass[prl,twocolumn,aps,showpacs,superscriptaddress]{revtex4-1}

\usepackage{graphicx}

\begin{document}

\title{Intrinsic Quantum Anomalous Hall Effect in the Kagome Lattice Cs$_2$LiMn$_3$F$_{12}$}

\author{Gang Xu, Biao Lian and Shou-Cheng Zhang}

\affiliation{Department of Physics, McCullough Building, Stanford University, Stanford, California 94305-4045, USA}

\date{\today}

\begin{abstract}
In a kagome lattice, the time reversal symmetry can be broken by a staggered magnetic flux emerging from ferromagnetic ordering and intrinsic spin-orbit coupling, leading to several well-separated nontrivial Chern bands and intrinsic quantum anomalous Hall effect. Based on this idea and \emph{ab~initio} calculations, we propose the realization of the intrinsic quantum anomalous Hall effect in the single layer Cs$_2$Mn$_3$F$_{12}$ kagome lattice and on the (001) surface of a Cs$_2$LiMn$_3$F$_{12}$ single crystal by modifying the carrier coverage on it, where the band gap is around 20 meV.
Moreover, a simplified tight binding model based on the in-plane $dd\sigma$ antibonding states is constructed to understand the topological band structures of the system.
\end{abstract}

\pacs{71.20.-b, 73.43.-f, 71.70.Ej, 73.20.At}
\maketitle
The quantum anomalous Hall (QAH) insulator is known as a topological state of matter in two dimensions (2D) with dissipationless chiral edge states protected by the electron band topology~\cite{haldane1988,Qi2006,Qi2008,Liu2008,Li2010,Yu2010,Xu2011,xiao2011,ruegg2011,Chang2013,wang2013a,wang2013b,zhang2014,Garrity2014,wang2014,kou2014,checkelsky2014,Xu2015,Chang2015}, which can be used to design new quantum devices such as the chiral interconnect~\cite{Zhang2012}. Haldane proposed a QAH model of a 2D honeycomb lattice with a staggered magnetic flux that averages out to zero~\cite{haldane1988}. Recently, the QAH effect has been theoretically predicted in magnetic topological insulators\cite{Qi2008,Liu2008,Li2010,Yu2010} and observed experimentally in Cr-doped (Bi,Sb)$_2$Te$_3$ (CBST) thin films at a low temperature around 30 mK~\cite{Chang2013}. However, such a low working temperature severely hinders the practical application of the QAH effect,
which is largely constrained by the spatial inhomogeneity induced by multiple random dopants in the system~\cite{Feng2015}.
For practical applications, to reach  a higher working temperature, it is important to search for intrinsic QAH materials without random doping
and increase the topologically nontrivial band gap, as well as the ferromagnetic (FM) Curie temperature ($T_c$).

\begin{figure}[htp]
\includegraphics[clip,width=2.4in,angle=270]{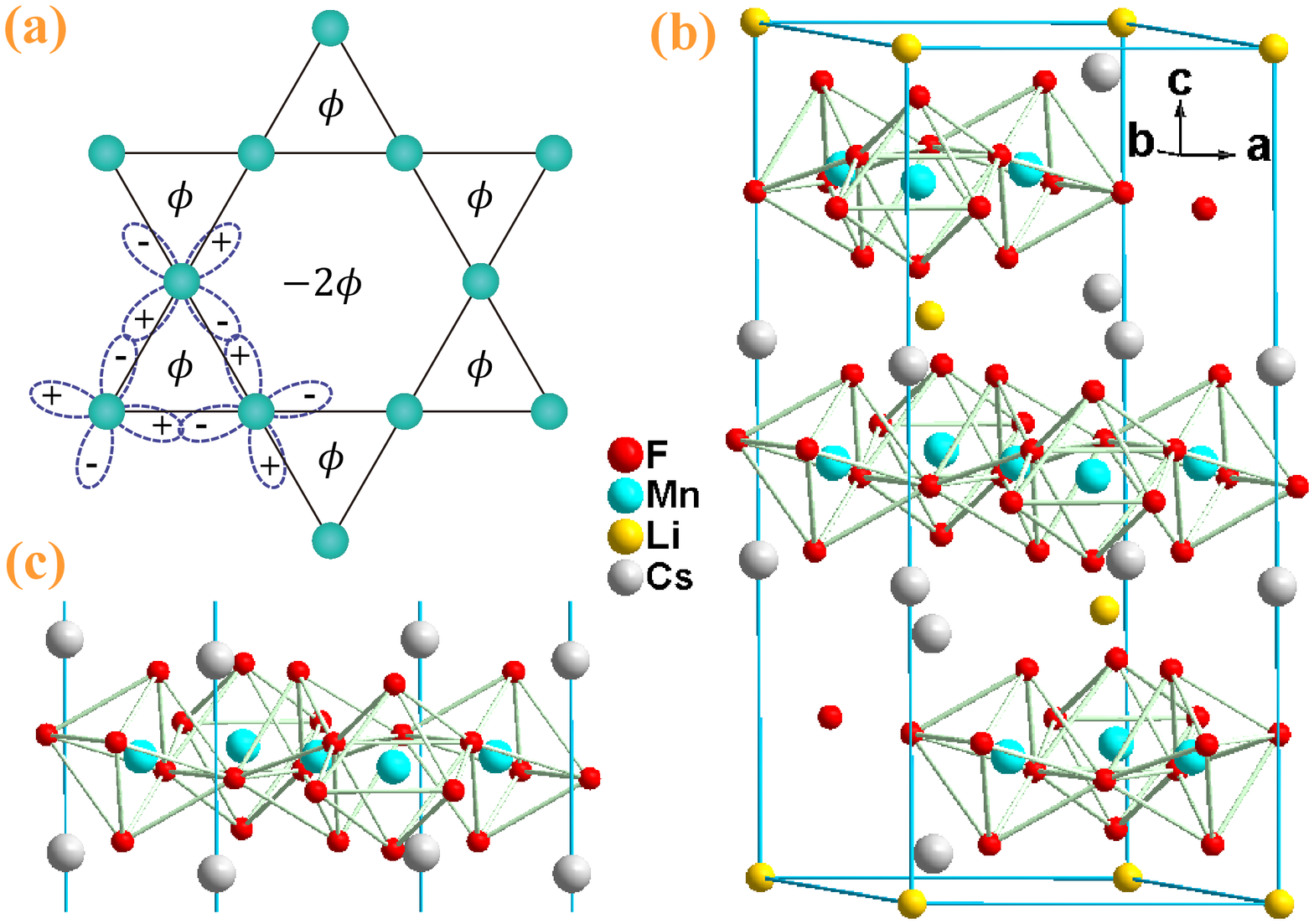}
\caption{(Color online) (a) Schematic of the Mn kagome lattice and the antibonding $d$ orbitals near the Fermi level. An effective staggered magnetic flux $\phi$ in the triangles and $-2\phi$ in the hexagons of the kagome lattice emerge due to the FM ordering and the intrinsic SOC. (b) The unit cell of Cs$_2$LiMn$_3$F$_{12}$ with space group $R\bar{3}$, which contains three primitive cells. (c), The crystal structure of single layer Cs$_2$Mn$_3$F$_{12}$.}\label{fig1}
\end{figure}

As is known, a honeycomb lattice with spin-orbit coupling (SOC) and FM ordering naturally realizes the Haldane model and the intrinsic QAH effect, and there are a lot of proposals for the QAH effect based on magnetic doped graphene (silicene) or graphene/FM (antiferromagnetic) heterostructures~\cite{Qiao2010,Zhang2012b,Ezawa2012,Qiao2014,Wang2015}. However, stoichiometric magnetic honeycomb compounds exist rarely in nature. On the other hand, the kagome lattice [Fig. 1(a)] shares similar topological physics with the honeycomb lattice~\cite{Ohgushi2000,Zhang2011}, and there are a number of magnetic materials adopting a kagome lattice structure~\cite{Englich1997,Shores2005,Wang2013c}.
With a staggered magnetic flux emerging from SOC and the FM ordering, the electronic bands of a 2D kagome lattice exhibit nontrivial Chern numbers, making it possible to realize the intrinsic QAH effect in such systems.
As an illustrative example, we propose here that the QAH effect can be realized in a single layer Cs$_2$Mn$_3$F$_{12}$ kagome lattice and on the (001) surface of a Cs$_2$LiMn$_3$F$_{12}$ single crystal
by modifying the surface carrier concentration so that the surface bands are at $2/3$ filling.
Compared to CBST, such~materials without any magnetic doping are easier to synthesize, and the much higher homogeneity and the large band gap around
20 meV enables a much higher working temperature.
A simplified tight binding (TB) model based on the in-plane $dd\sigma$ antibonding states is constructed to understand the topological physics of the system. While most studies of the kagome lattice are focused on the flat band physics at $1/3$ filling~\cite{Tang2011,Venderbos2011,Nishimoto2010,Green2010,Liu2010b,Nakata2012}, the system considered here is at $2/3$ filling with a low energy theory governed by Dirac fermions at $\bar{K}$ and $\bar{K'}$ points similar to that in the Haldane model. This class of materials opens a new direction for the exploration of the topological states of matter.

Experimentally, the Cs$_2$LiMn$_3$F$_{12}$ single crystal is synthesized within the $R\bar{3}$ space group (point group $S_6$)~\cite{Englich1997},
where the Mn atoms are surrounded by a corner-sharing octahedron constructed by F atoms and form perfect kagome lattices layer by layer as shown in Fig. 1(b) and Fig. 1(a). Each [Mn$_3$F$_{12}$] kagome layer is sandwiched by two layers of Cs atoms and connected by Li atoms between [Cs$_2$Mn$_3$F$_{12}$] layers along c axis (z direction). We note that the [MnF$_{6}$] octahedra are strongly elongated by the Jahn-Teller distortion with ratios of long (within the layer) to short axes (out of the layer) of about 1.18 and form an antiferrodistortive Jahn-Teller ordering. This is very similar to the FM manganese fluoride CsMnF$_{4}$, where the antiferrodistortive ordering of Jahn-Teller elongated [MnF$_{6}$] octahedra with the ratio 1.17 makes the asymmetric $\sigma$ superexchange interactions between the empty $d_{z^2}$ orbitals and half-filled $d_{x^2-y^2}$ orbitals of Mn$^{3+}$ favor the planar FM coupling with $T_c$ = 19 $K$~\cite{Massa1980}. Because of such a large Jahn-Teller distortion, it is the trigonal rather than the octahedral field that dominates the splitting of the $3d$ orbital of Mn as shown below.

Our $ab ~initio$ calculations are carried out by the density functional method~\cite{Hohenberg1964,Kohn1965} based on the plane-wave ultrasoft pseudopotential approximation~\cite{Vanderbilt1990} with the generalized gradient approximation of the Perdew-Burke-Ernzerhof type for the exchange-correlation potential~\cite{Perdew1996}, as implemented in the BSTATE (Beijing Simulation Tool of Atomic Technology) package~\cite{Fang2002}.
Fully optimized lattice constants $a$ = 7.5698 \AA, $c$ = 17.4362 \AA ~with the force smaller than 0.01 eV/\AA ~are used in all calculations, which are very close to the experimental measurements $a$ = 7.440 \AA, $c$ = 17.267 \AA. A slab constructed by three layers of Cs$_2$LiMn$_3$F$_{12}$ with more than 18 \AA of vacuum are used for the surface study, and the virtual crystal approximation (VCA)~\cite{Nordheim1931,Bellaiche2000} is used to simulate the carrier concentration on the surface. The cutoff energy for wave function expansion is set to 400 eV, and $8\times8\times8$ and $8\times8\times2$ k meshes are used for the bulk and slab self-consistent calculations, respectively. SOC effect is considered consistently in the calculations. All results except VCA calculations are well reproduced by VASP~\cite{Kresse1993,Kresse1996}.

Our calculations conclude that the $3d$ states of Mn$^{3+}$ are strongly spin polarized and Cs$_2$LiMn$_3$F$_{12}$ is a FM insulator with magnetic moment of 4.0 $\mu_B$/Mn. The calculated total energy of the FM solution is 6.71 eV/f.u. lower than the nonmagnetic solution and 1.57 eV/f.u. lower than the all-in-all-out antiferromagnetic (AFM) solution, which confirms that the antiferrodistortive ordering of Jahn-Teller elongated [MnF$_{6}$] octahedra favors an in-plane FM exchange coupling between the magnetic moment of Mn. This is quite different from the spin liquid system herbertsmithite ZnCu$_3$(OH)$_6$Cl$_2$, where the Cu atoms also form the layered kagome lattice but the magnetic moment of Cu favors an in-plane AFM exchange coupling ~\cite{Shores2005,Jeschke2013}.

\begin{figure}[tbp]
\includegraphics[clip,scale=0.41, angle=270]{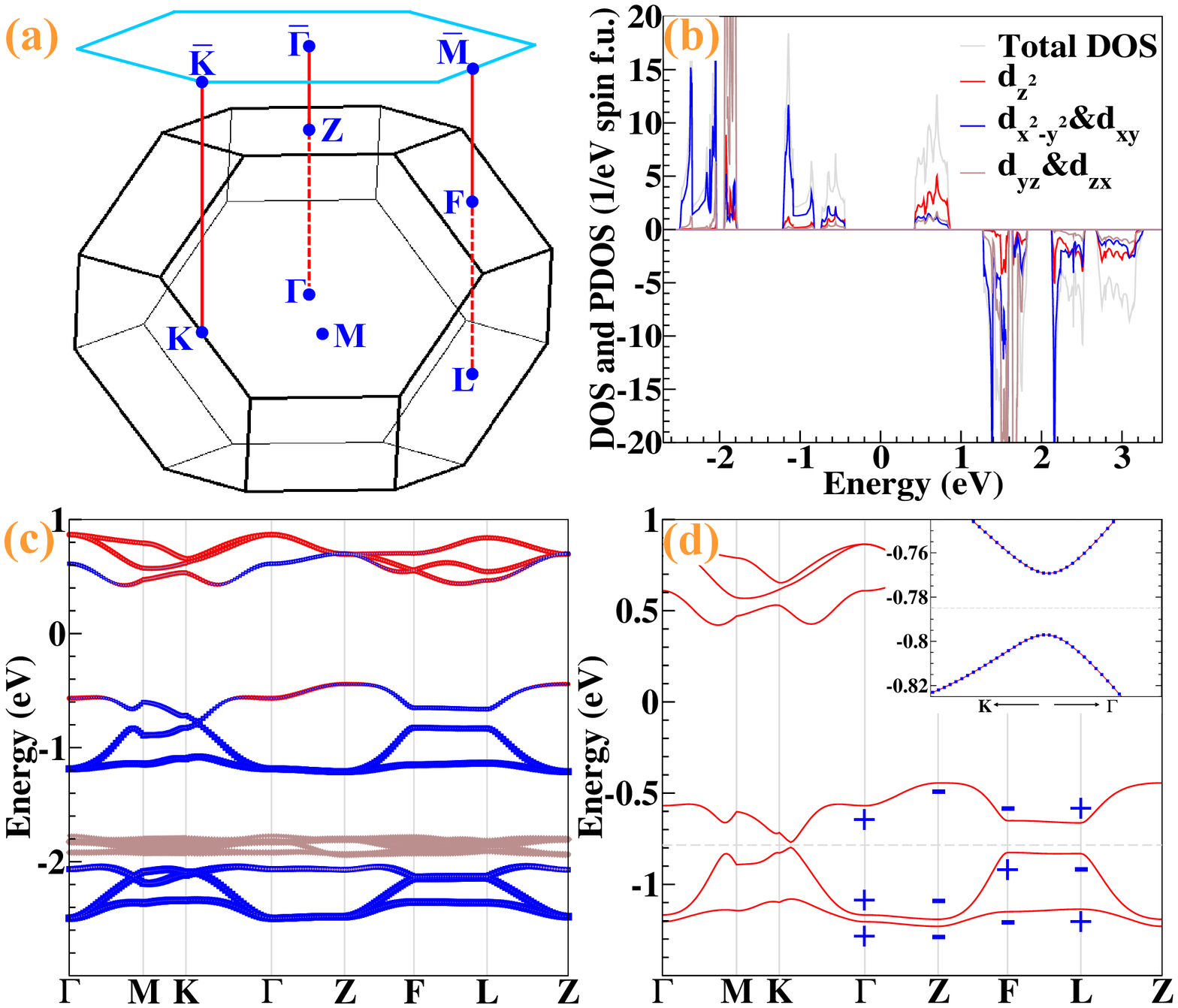}
\caption{(Color online) (a) The first Brillouin zones of Cs$_2$LiMn$_3$F$_{12}$ with space group $R\bar{3}$. The four types inversion-invariant points are $\Gamma(0,0,0)$, $Z$($\pi, \pi, \pi$), $F$($\pi,\pi$,0) and $L$($\pi$,0,0). The blue hexagon shows the 2D Brillouin zones for the single layer Cs$_2$Mn$_3$F$_{12}$ and slab, in which the high-symmetry $k$ points $\bar{\Gamma}$, $\bar{M}$ and $\bar{K}$ are labeled.
(b) Spin resolved DOS and PDOS of the bulk Cs$_2$LiMn$_3$F$_{12}$. (c) FM band structures of Cs$_2$LiMn$_3$F$_{12}$ without SOC,
in which red circles, blue squares and brown diamonds denote the projections on the $d_{z^2}$, $d_{x^2-y^2}\&d_{xy}$, and $d_{yz}\&d_{zx}$ orbitals of Mn, respectively.
(d) FM band structures of Cs$_2$LiMn$_3$F$_{12}$ with SOC, in which the parities of the topmost three valence bands at eight inversion-symmetric points are marked. The inset is the enlargement around the $K$ point to show the fully opened gap. All Fermi levels are defined at 0 eV.}
\end{figure}

In Fig. 2(b), we show the spin resolved density of states (DOS) and projected density of states (PDOS) on Mn for the FM solution, in which the states around the Fermi level (0 eV) are mostly contributed by Mn-$3d$ up spin states, while all the down spin Mn-$3d$ states are pushed above more than 1 eV away due to the large Hund's rule coupling. With the trigonal field effect taken into account, the up spin Mn-$3d$ states roughly split into three groups: $d_{z^2}$, $d_{x^2-y^2}\&d_{xy}$, and $d_{yz}\&d_{zx}$, as represented by the red circles, blue squares and brown diamonds in Fig. 2(c), respectively.
The $d_{x^2-y^2}\&d_{xy}$ group further splits into the bonding states $d^{\dagger}_{x^2-y^2}\&d^{\dagger}_{xy}$ and the antibonding states $d^{\ast}_{x^2-y^2}\&d^{\ast}_{xy}$ due to the in-plane Mn-Mn $dd\sigma$ interaction. As a result, the three valence states and three conduction states are $d^{\ast}_{x^2-y^2}\&d^{\ast}_{xy}\uparrow$ and $d_{z^2}\uparrow$, respectively, as shown in Fig. 2(b) and Fig 2(c). In Fig. 2(c), there are two important features to be pointed out. (1) The dispersion along the $z$ direction is very weak (see the bands along $\Gamma-Z$, $F-L$), indicating that Cs$_2$LiMn$_3$F$_{12}$ has a weak interlayer hopping and is therefore a good layer by layer material. (2) In the absence of SOC, the highest two valence bands cross with each other near the $K$ points, and form six  Weyl points protected by the crystal symmetry. Once the SOC is included, an insulating gap opens at the Weyl points, resulting in a topological nontrivial phase if the Fermi level is tuned to the gray dashed line as shown in Fig 2(d), \emph{i.e.}, one hole doped per primitive cell. The parities of the Bloch wave functions of the topmost three valence bands at eight inversion-symmetric points $\Gamma(0, 0, 0)$, $Z$($\pi, \pi, \pi$), $3F$($\pi, \pi$, 0), $3L$($\pi$, 0, 0) are calculated and labeled in Fig. 2(d). Based on the parity analysis, if one electron per primitive cell is reduced from the system, the occupied bands below the Fermi level [the gray dashed line in Fig. 2(d)] will have weak topological numbers $(\nu_0, \nu_1, \nu_2, \nu_3)$ = (0, 1, 1, 1), which indicates the system can be viewed as many 2D QAH insulator layers weakly stacked along (111), \emph{i.e.}, the $z$ direction.

However, it is too hard to reduce one electron directly from the 3D bulk of the material. Instead, it is much easier to modify the carrier concentration on the surface or in the 2D limit (single layer) for such kinds of layered materials~\cite{Weng2011,Yao2015}. Experimentally, single layer Cs$_2$Mn$_3$F$_{12}$ as shown in Fig. 1(c) can be cleaved from the Cs$_2$LiMn$_3$F$_{12}$ single crystal. After the cleaving, two kinds of terminations, with or without the outermost Li layer, may be realized. Nevertheless, since the outermost Li atoms are highly mobile and active, their concentration can be tuned depending on the experimental conditions. For example, we calculated the total energy of the fluorine molecule F$_2$, LiF, the single layer Li$_2$Cs$_2$Mn$_3$F$_{12}$ and Cs$_2$Mn$_3$F$_{12}$, and found that $E$(Cs$_2$Mn$_3$F$_{12}$)+$2E$(LiF) is 1.78eV lower than $E$(Li$_2$Cs$_2$Mn$_3$F$_{12}$)+$E$(F$_2$), which clearly shows that the outermost Li atoms can be removed in the F$_2$-rich environment. On the other hand, because the main function of the Li atoms in Cs$_2$LiMn$_3$F$_{12}$ is to provide carriers, removing them changes little the electronic structures and the stability of the lattice. To test this, we have optimized the structure of single layer Cs$_2$Mn$_3$F$_{12}$ with a vacuum region thicker than 15 \AA, and have confirmed that all of its structure parameters are nearly the same as those of the bulk Cs$_2$LiMn$_3$F$_{12}$. Furthermore,
our calculations show that single layer Cs$_2$Mn$_3$F$_{12}$ also favors a FM solution with a total magnetic moment 3.6 $\mu_B$/Mn, which is 5.85 eV/(single layer) lower than the nonmagnetic solution and 1.58 eV/(single layer) lower than the all-in-all-out AFM solution. The calculated FM band structures of single layer Cs$_2$Mn$_3$F$_{12}$ without (blue) and with (red) SOC are shown in Fig. 3(b), where the three bands near the Fermi level are still mainly from the $d^{\ast}_{x^2-y^2}\&d^{\ast}_{xy}\uparrow$ orbitals.

\begin{figure}[tbp]
\includegraphics[clip,scale=0.39,angle=270]{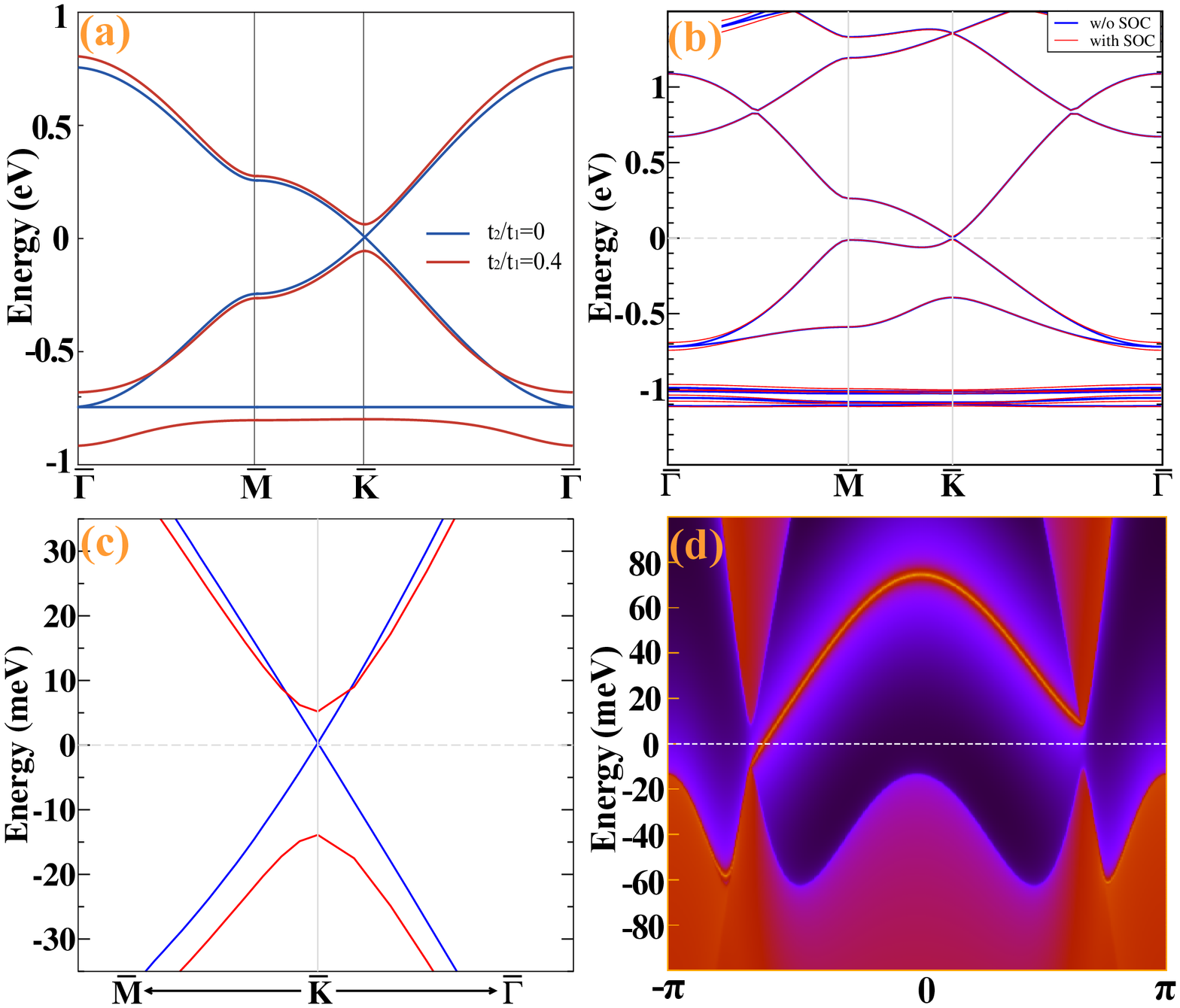}
\caption{(Color online) (a) TB band structures with $t_1$ = 0.25 eV, $t_{2}$ = 0 (blue) and $t_2=0.4t_1$ (red), respectively.
(b) FM band structures of single layer Cs$_2$Mn$_3$F$_{12}$ with (red) and without (blue) SOC calculations.
(c) Enlargement of (b) around the $K$ point to show the Dirac point (without SOC) and topologically nontrivial gap (with SOC) clearly.
(d) Calculated edge state for the semi-infinite boundary of the 2D kagome lattice Cs$_2$Mn$_3$F$_{12}$. All Fermi levels are defined at 0 eV.
  }
\end{figure}

To understand the topological properties of single layer Cs$_2$Mn$_3$F$_{12}$, we construct here a TB model of the kagome lattice based on the three antibonding $d^{\ast}_{x^2-y^2}\&d^{\ast}_{xy}$ states near the Fermi level. The corresponding antibonding basis wave functions of the three Mn atoms in one unit cell are shown in Fig. 1(a), which are related to each other by $\pm2\pi/3$ rotations. Meanwhile, the electron spins are polarized due to the FM ordering. In the presence of an intrinsic SOC, the nearest neighbor spin polarized TB model takes the form
\begin{equation}
\mathcal{H}=\sum_{\langle ij\rangle}(t_1+it_2\nu_{ij})c^\dag_{i\uparrow}c_{j\uparrow}\ ,
\end{equation}
where $c_{i\alpha}$ ($c^\dag_{i\alpha}$) is the annihilation (creation) operator of an electron of spin $\alpha$ ($\alpha=\uparrow, \downarrow$) on site $i$, $t_1$, $t_2$ are real, and $\nu_{ij}=+1(-1)$ if $j\rightarrow i$ is hopping counterclockwise (clockwise) about the triangle containing site $i$ and site $j$. $t_1>0$ represents the direct hopping amplitude, while $t_2$ comes from the intrinsic SOC term $\sum_{\langle ij\rangle\alpha\beta}it_2\nu_{ij}c^\dag_{i\alpha}s^z_{\alpha\beta}c_{j\beta}$, where $s^z$ is the Pauli matrix for the electron spin~\cite{Tang2011,kane2005a}. As a result, there is a magnetic flux $\phi=3\arctan(t_2/t_1)$ in the triangles and $-2\phi$ in the hexagons of the kagome lattice, respectively, as shown in Fig. 1(a). We note that such a staggered magnetic flux averages out to zero and preserves the inversion symmetry. This is quite different from another class of kagome lattice models relying on FM ordering and Rashba SOC that breaks the inversion symmetry~\cite{Zhang2011}.

The physics of this TB model is quite similar to that of the Haldane model. The model consists of two dispersive upper bands $|u_1\rangle$, $|u_2\rangle$ and a lower flat band $|u_3\rangle$. When $t_2=0$ (the flux $\phi=0$), the upper band $|u_1\rangle$ and the middle band $|u_2\rangle$ touch linearly at the $\bar{K}$ and $\bar{K'}$ points, forming two Dirac cones with opposite helicities, while the middle band $|u_2\rangle$ and the lower flat band $|u_3\rangle$ touch quadratically at the $\Gamma$ point. When $t_2>0$ (the flux $0<\phi<\pi$), the three bands $|u_1\rangle, |u_2\rangle, |u_3\rangle$ are detached from each other in energies, and carry Chern numbers $-1,0,+1$, respectively. To be explicit, we have plotted the band structures for the TB model with $t_2/t_1=0$ and $t_2/t_1=0.4$, respectively in Fig. 3(a), where we have set $t_1 = 0.25$ eV. We note that the QAH effect can be realized at both 1/3 filling and 2/3 filling. While most existing theoretical works are focused on 1/3 filling flat band physics and the related fractional QAH effect, it is hard to obtain a full insulating band gap at 1/3
filling in the presence of hopping beyond nearest neighbors in real materials [Fig. 3(b)]. On the contrary, it is much easier to open an insulating band gap at 2/3 filling as shown in Fig. 3(c) and Fig. 4.

For small $t_2$, we can write down an effective Hamiltonian in the continuous limit:
\begin{equation}
\mathcal{H}=-i\hbar v_F\psi^\dag(\sigma_x\tau_z\partial_x+\sigma_y\partial_y)\psi+m_{soc}\psi^\dag\sigma_z\tau_z\psi\ ,
\end{equation}
where $\psi$ is a four component wave function, $\sigma_{x,y,z}$ are Pauli matrices describing the lattice pseudospin, and $\tau_z=\pm1$ describes states at the $\bar{K}$ and $\bar{K'}$ points, respectively. The mass term $m_{soc}$ is given by $m_{soc} = 2t_2/3$. Such a low energy Hamiltonian has exactly the same form as that of the Haldane model~\cite{haldane1988}, which makes it much easier to understand the topological phase transition. When $m_{soc} > 0$, each of the two Dirac cones contributes a $\pi$ Berry phase flux, leading to a total Chern number $C=+1$, while, when $m_{soc} < 0$, both of the Berry phase fluxes change sign and the total Chern number becomes $C=-1$.

As shown in Fig. 3(b) and Fig. 3(c), our calculations confirm that single layer Cs$_2$Mn$_3$F$_{12}$ without SOC is a gapless semimetal with the Fermi level exactly crossing two Dirac points at $\bar{K}$ and $\bar{K'}$, similar to that in graphene. When SOC is introduced into the system, an insulating gap of 20 meV is opened and the system becomes a QAH insulator with Chern number $C = 1$, as demonstrated by our TB model. We have carried out the calculations of edge states by constructing the Green's functions~\cite{Sancho1985} for the semi-infinite boundary based on the maximally Localized Wannier function method~\cite{Marzari1997,Souza2001}. The results are shown in Fig. 3(d), in which one topologically protected chiral edge state is present in between the valence band and the conduction band, clearly indicating the system has Chern number $C = 1$.

\begin{figure}[tbp]
\includegraphics[clip,scale=0.361,angle=270]{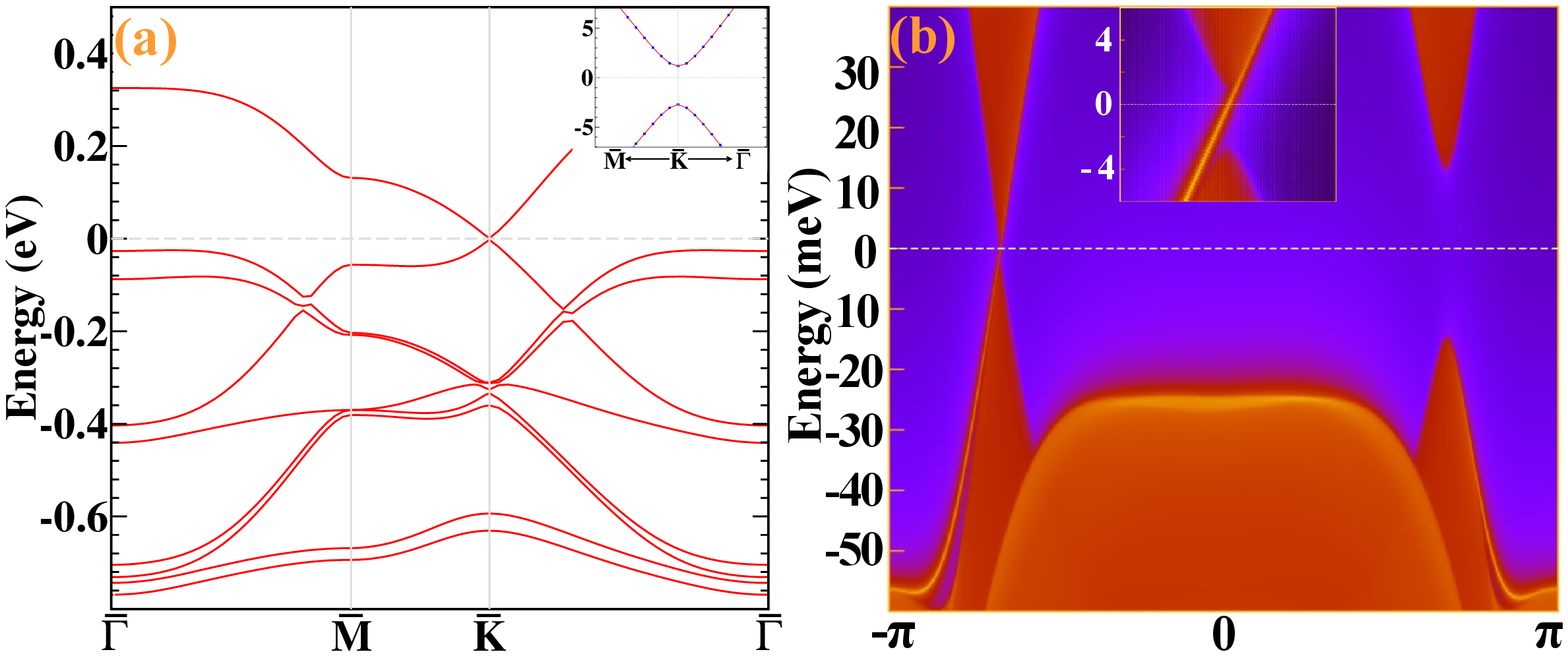}
\caption{(Color online) (a) FM band structures of the slab constructed by three layers of Cs$_2$LiMn$_3$F$_{12}$ with SOC.
The inset is the enlargement around $K$ point to show the topologically nontrivial gap (4 meV) clearly.
(b) Calculated edge state for the slab constructed by three layers of Cs$_2$LiMn$_3$F$_{12}$.
The inset is the enlargement around the Fermi level to show the insulating gap (4 meV) and chiral edge state clearly.
All Fermi levels are defined at 0 eV.
  }
\end{figure}

Another proposal to achieve the required Fermi filling to realize the QAH effect in Cs$_2$LiMn$_3$F$_{12}$ is to modify the carrier concentration on the surface. The layered crystal structure guarantees that the sample can be easily cleaved, and the carrier concentration on the surface can be tuned by the coverage of the topmost Li and Cs atoms. By passivating the lower surface, we optimize the structure of the upper surface (slab) without the topmost Li layer. Our calculations confirm that the structure variation of the surface layer Cs$_2$Mn$_3$F$_{12}$ is negligible, and the hole only goes into the surface layer, keeping all the other bulk layers insulating as shown in Fig. 4(a). The calculated magnetic moments are 3.6 $\mu_B$/Mn for the surface layer, and 4.0 $\mu_B$/Mn for the bulk layers. As shown in Fig. 4(a), there are only two bands coming from the $d^{\ast}_{x^2-y^2}\&d^{\ast}_{xy}\uparrow$ states of the topmost Mn layer crossing the Fermi level, while all the other $d^{\ast}_{x^2-y^2}\&d^{\ast}_{xy}\uparrow$ bands of the bulk Mn layers are fully occupied. Therefore, the surface layer of Mn behaves like a pure 2D kagome lattice, in which the QAH effect can be achieved with suitable carrier concentration. The band structures for the slab with SOC exhibits an insulating gap of about 4 meV on the surface, as shown in the inset of Fig. 4(a). The calculated chiral edge state of the slab is shown in Fig. 4(b), showing that the QAH effect with Chern number $C = 1$ is realized on the surface of the slab.

In conclusion, we have studied the possibility of realizing the intrinsic QAH effect in materials with a ferromagnetic kagome lattice structure, and have predicted that Cs$_2$LiMn$_3$F$_{12}$ is such a good candidate. We constructed a simplified TB model with ferromagnetic ordering and SOC based on the in-plane $dd\sigma$ antibonding states, and demonstrated the realization of the QAH effect at $2/3$ filling.
Based on \emph{ab initio} calculations, we showed that the single layer Cs$_2$Mn$_3$F$_{12}$ ferromagnetic kagome lattice has exactly $2/3$ Fermi filling, and realizes an intrinsic QAH insulator with a band gap of around $20$ meV. Such nondoped materials are much easier to synthesize and much more homogeneous than CBST, therefore enables a much higher working temperature than that in CBST. Further, our calculations suggested that a similar Fermi filling and QAH effect can also be achieved on the (001) surface layer of a Cs$_2$LiMn$_3$F$_{12}$ single crystal via a modification of the surface carrier coverage.

We acknowledge valuable discussions with Xiao-Liang Qi.
This work is supported by the U.S. Department of Energy, Office of Basic Energy Sciences, Division of Materials Sciences and Engineering under Contract No.~DE-AC02-76SF00515, by FAME, one of six centers of STARnet, a Semiconductor Research Corporation program sponsored by MARCO and DARPA.


\begin{thebibliography}{54}
\makeatletter
\providecommand \@ifxundefined [1]{%
 \@ifx{#1\undefined}
}%
\providecommand \@ifnum [1]{%
 \ifnum #1\expandafter \@firstoftwo
 \else \expandafter \@secondoftwo
 \fi
}%
\providecommand \@ifx [1]{%
 \ifx #1\expandafter \@firstoftwo
 \else \expandafter \@secondoftwo
 \fi
}%
\providecommand \natexlab [1]{#1}%
\providecommand \enquote  [1]{``#1''}%
\providecommand \bibnamefont  [1]{#1}%
\providecommand \bibfnamefont [1]{#1}%
\providecommand \citenamefont [1]{#1}%
\providecommand \href@noop [0]{\@secondoftwo}%
\providecommand \href [0]{\begingroup \@sanitize@url \@href}%
\providecommand \@href[1]{\@@startlink{#1}\@@href}%
\providecommand \@@href[1]{\endgroup#1\@@endlink}%
\providecommand \@sanitize@url [0]{\catcode `\\12\catcode `\$12\catcode
  `\&12\catcode `\#12\catcode `\^12\catcode `\_12\catcode `\%12\relax}%
\providecommand \@@startlink[1]{}%
\providecommand \@@endlink[0]{}%
\providecommand \url  [0]{\begingroup\@sanitize@url \@url }%
\providecommand \@url [1]{\endgroup\@href {#1}{\urlprefix }}%
\providecommand \urlprefix  [0]{URL }%
\providecommand \Eprint [0]{\href }%
\providecommand \doibase [0]{http://dx.doi.org/}%
\providecommand \selectlanguage [0]{\@gobble}%
\providecommand \bibinfo  [0]{\@secondoftwo}%
\providecommand \bibfield  [0]{\@secondoftwo}%
\providecommand \translation [1]{[#1]}%
\providecommand \BibitemOpen [0]{}%
\providecommand \bibitemStop [0]{}%
\providecommand \bibitemNoStop [0]{.\EOS\space}%
\providecommand \EOS [0]{\spacefactor3000\relax}%
\providecommand \BibitemShut  [1]{\csname bibitem#1\endcsname}%
\let\auto@bib@innerbib\@empty
\bibitem [{\citenamefont {Haldane}(1988)}]{haldane1988}%
  \BibitemOpen
  \bibfield  {author} {\bibinfo {author} {\bibfnamefont {F.~D.~M.}\
  \bibnamefont {Haldane}},\ }\href@noop {} {\bibfield  {journal} {\bibinfo
  {journal} {Phys. Rev. Lett.}\ }\textbf {\bibinfo {volume} {61}},\ \bibinfo
  {pages} {2015} (\bibinfo {year} {1988})}\BibitemShut {NoStop}%
\bibitem [{\citenamefont {Qi}\ \emph {et~al.}(2006)\citenamefont {Qi},
  \citenamefont {Wu},\ and\ \citenamefont {Zhang}}]{Qi2006}%
  \BibitemOpen
  \bibfield  {author} {\bibinfo {author} {\bibfnamefont {X.~L.}\ \bibnamefont
  {Qi}}, \bibinfo {author} {\bibfnamefont {Y.~S.}\ \bibnamefont {Wu}}, \ and\
  \bibinfo {author} {\bibfnamefont {S.~C.}\ \bibnamefont {Zhang}},\ }\href@noop
  {} {\bibfield  {journal} {\bibinfo  {journal} {Phys. Rev. B}\ }\textbf
  {\bibinfo {volume} {74}},\ \bibinfo {pages} {045125} (\bibinfo {year}
  {2006})}\BibitemShut {NoStop}%
\bibitem [{\citenamefont {Qi}\ \emph {et~al.}(2008)\citenamefont {Qi},
  \citenamefont {Hughes},\ and\ \citenamefont {Zhang}}]{Qi2008}%
  \BibitemOpen
  \bibfield  {author} {\bibinfo {author} {\bibfnamefont {X.-L.}\ \bibnamefont
  {Qi}}, \bibinfo {author} {\bibfnamefont {T. L.}~\bibnamefont {Hughes}}, \ and\
  \bibinfo {author} {\bibfnamefont {S.-C.}\ \bibnamefont {Zhang}},\ }\href@noop
  {} {\bibfield  {journal} {\bibinfo  {journal} {Phys. Rev. B}\ }\textbf
  {\bibinfo {volume} {78}},\ \bibinfo {pages} {195424} (\bibinfo {year}
  {2008})}\BibitemShut {NoStop}%
\bibitem [{\citenamefont {Liu}\ \emph {et~al.}(2008)\citenamefont {Liu},
  \citenamefont {Qi}, \citenamefont {Dai}, \citenamefont {Fang},\ and\
  \citenamefont {Zhang}}]{Liu2008}%
  \BibitemOpen
  \bibfield  {author} {\bibinfo {author} {\bibfnamefont {C.-X.}\ \bibnamefont
  {Liu}}, \bibinfo {author} {\bibfnamefont {X.-L.}\ \bibnamefont {Qi}},
  \bibinfo {author} {\bibfnamefont {X.}~\bibnamefont {Dai}}, \bibinfo {author}
  {\bibfnamefont {Z.}~\bibnamefont {Fang}}, \ and\ \bibinfo {author}
  {\bibfnamefont {S.-C.}\ \bibnamefont {Zhang}},\ }\href@noop {} {\bibfield
  {journal} {\bibinfo  {journal} {Phys. Rev. Lett.}\ }\textbf {\bibinfo
  {volume} {101}},\ \bibinfo {pages} {146802} (\bibinfo {year}
  {2008})}\BibitemShut {NoStop}%
\bibitem [{\citenamefont {Li}\ \emph {et~al.}(2010)\citenamefont {Li},
  \citenamefont {Wang}, \citenamefont {Qi},\ and\ \citenamefont
  {Zhang}}]{Li2010}%
  \BibitemOpen
  \bibfield  {author} {\bibinfo {author} {\bibfnamefont {R.}~\bibnamefont
  {Li}}, \bibinfo {author} {\bibfnamefont {J.}~\bibnamefont {Wang}}, \bibinfo
  {author} {\bibfnamefont {X.~L.}\ \bibnamefont {Qi}}, \ and\ \bibinfo {author}
  {\bibfnamefont {S.~C.}\ \bibnamefont {Zhang}},\ }\href@noop {} {\bibfield
  {journal} {\bibinfo  {journal} {Nature Phys.}\ }\textbf {\bibinfo {volume}
  {6}},\ \bibinfo {pages} {284} (\bibinfo {year} {2010})}\BibitemShut {NoStop}%
\bibitem [{\citenamefont {Yu}\ \emph {et~al.}(2010)\citenamefont {Yu},
  \citenamefont {Zhang}, \citenamefont {Zhang}, \citenamefont {Zhang},
  \citenamefont {Dai},\ and\ \citenamefont {Fang}}]{Yu2010}%
  \BibitemOpen
  \bibfield  {author} {\bibinfo {author} {\bibfnamefont {R.}~\bibnamefont
  {Yu}}, \bibinfo {author} {\bibfnamefont {W.}~\bibnamefont {Zhang}}, \bibinfo
  {author} {\bibfnamefont {H.~J.}\ \bibnamefont {Zhang}}, \bibinfo {author}
  {\bibfnamefont {S.~C.}\ \bibnamefont {Zhang}}, \bibinfo {author}
  {\bibfnamefont {X.}~\bibnamefont {Dai}}, \ and\ \bibinfo {author}
  {\bibfnamefont {Z.}~\bibnamefont {Fang}},\ }\href@noop {} {\bibfield
  {journal} {\bibinfo  {journal} {Science}\ }\textbf {\bibinfo {volume}
  {329}},\ \bibinfo {pages} {61} (\bibinfo {year} {2010})}\BibitemShut
  {NoStop}%
\bibitem [{\citenamefont {Xu}\ \emph {et~al.}(2011)\citenamefont {Xu},
  \citenamefont {Weng}, \citenamefont {Wang}, \citenamefont {Dai},\ and\
  \citenamefont {Fang}}]{Xu2011}%
  \BibitemOpen
  \bibfield  {author} {\bibinfo {author} {\bibfnamefont {G.}~\bibnamefont
  {Xu}}, \bibinfo {author} {\bibfnamefont {H.}~\bibnamefont {Weng}}, \bibinfo
  {author} {\bibfnamefont {Z.}~\bibnamefont {Wang}}, \bibinfo {author}
  {\bibfnamefont {X.}~\bibnamefont {Dai}}, \ and\ \bibinfo {author}
  {\bibfnamefont {Z.}~\bibnamefont {Fang}},\ }\href {\doibase
  10.1103/PhysRevLett.107.186806} {\bibfield  {journal} {\bibinfo  {journal}
  {Phys. Rev. Lett.}\ }\textbf {\bibinfo {volume} {107}},\ \bibinfo {pages}
  {186806} (\bibinfo {year} {2011})}\BibitemShut {NoStop}%
\bibitem [{\citenamefont {Xiao}\ \emph {et~al.}(2011)\citenamefont {Xiao},
  \citenamefont {Zhu}, \citenamefont {Ran}, \citenamefont {Nagaosa},\ and\
  \citenamefont {Okamoto}}]{xiao2011}%
  \BibitemOpen
  \bibfield  {author} {\bibinfo {author} {\bibfnamefont {D.}~\bibnamefont
  {Xiao}}, \bibinfo {author} {\bibfnamefont {W.}~\bibnamefont {Zhu}}, \bibinfo
  {author} {\bibfnamefont {Y.}~\bibnamefont {Ran}}, \bibinfo {author}
  {\bibfnamefont {N.}~\bibnamefont {Nagaosa}}, \ and\ \bibinfo {author}
  {\bibfnamefont {S.}~\bibnamefont {Okamoto}},\ }\href@noop {} {\bibfield
  {journal} {\bibinfo  {journal} {Nat. Commun.}\ }\textbf {\bibinfo {volume}
  {2}},\ \bibinfo {pages} {596} (\bibinfo {year} {2011})}\BibitemShut {NoStop}%
\bibitem [{\citenamefont {R\"uegg}\ and\ \citenamefont
  {Fiete}(2011)}]{ruegg2011}%
  \BibitemOpen
  \bibfield  {author} {\bibinfo {author} {\bibfnamefont {A.}~\bibnamefont
  {R\"uegg}}\ and\ \bibinfo {author} {\bibfnamefont {G.~A.}\ \bibnamefont
  {Fiete}},\ }\href {\doibase 10.1103/PhysRevB.84.201103} {\bibfield  {journal}
  {\bibinfo  {journal} {Phys. Rev. B}\ }\textbf {\bibinfo {volume} {84}},\
  \bibinfo {pages} {201103} (\bibinfo {year} {2011})}\BibitemShut {NoStop}%
\bibitem [{\citenamefont {Chang}\ \emph {et~al.}(2013)\citenamefont {Chang},
  \citenamefont {Zhang}, \citenamefont {Feng}, \citenamefont {Shen},
  \citenamefont {Zhang}, \citenamefont {Guo}, \citenamefont {Li}, \citenamefont
  {Ou}, \citenamefont {Wei}, \citenamefont {Wang}, \citenamefont {Ji},
  \citenamefont {Feng}, \citenamefont {Ji}, \citenamefont {Chen}, \citenamefont
  {Jia}, \citenamefont {Dai}, \citenamefont {Fang}, \citenamefont {Zhang},
  \citenamefont {He}, \citenamefont {Wang}, \citenamefont {Lu}, \citenamefont
  {Ma},\ and\ \citenamefont {Xue}}]{Chang2013}%
  \BibitemOpen
  \bibfield  {author} {\bibinfo {author} {\bibfnamefont {C.-Z.}\ \bibnamefont
  {Chang}}, \bibinfo {author} {\bibfnamefont {J.}~\bibnamefont {Zhang}},
  \bibinfo {author} {\bibfnamefont {X.}~\bibnamefont {Feng}}, \bibinfo {author}
  {\bibfnamefont {J.}~\bibnamefont {Shen}}, \bibinfo {author} {\bibfnamefont
  {Z.}~\bibnamefont {Zhang}}, \bibinfo {author} {\bibfnamefont
  {M.}~\bibnamefont {Guo}}, \bibinfo {author} {\bibfnamefont {K.}~\bibnamefont
  {Li}}, \bibinfo {author} {\bibfnamefont {Y.}~\bibnamefont {Ou}}, \bibinfo
  {author} {\bibfnamefont {P.}~\bibnamefont {Wei}}, \bibinfo {author}
  {\bibfnamefont {L.-L.}\ \bibnamefont {Wang}}, \bibinfo {author}
  {\bibfnamefont {Z.-Q.}\ \bibnamefont {Ji}}, \bibinfo {author} {\bibfnamefont
  {Y.}~\bibnamefont {Feng}}, \bibinfo {author} {\bibfnamefont {S.}~\bibnamefont
  {Ji}}, \bibinfo {author} {\bibfnamefont {X.}~\bibnamefont {Chen}}, \bibinfo
  {author} {\bibfnamefont {J.}~\bibnamefont {Jia}}, \bibinfo {author}
  {\bibfnamefont {X.}~\bibnamefont {Dai}}, \bibinfo {author} {\bibfnamefont
  {Z.}~\bibnamefont {Fang}}, \bibinfo {author} {\bibfnamefont {S.-C.}\
  \bibnamefont {Zhang}}, \bibinfo {author} {\bibfnamefont {K.}~\bibnamefont
  {He}}, \bibinfo {author} {\bibfnamefont {Y.}~\bibnamefont {Wang}}, \bibinfo
  {author} {\bibfnamefont {L.}~\bibnamefont {Lu}}, \bibinfo {author}
  {\bibfnamefont {X.-C.}\ \bibnamefont {Ma}}, \ and\ \bibinfo {author}
  {\bibfnamefont {Q.-K.}\ \bibnamefont {Xue}},\ }\href {\doibase
  10.1126/science.1234414} {\bibfield  {journal} {\bibinfo  {journal}
  {Science}\ }\textbf {\bibinfo {volume} {340}},\ \bibinfo {pages} {167}
  (\bibinfo {year} {2013})}\BibitemShut {NoStop}%
\bibitem [{\citenamefont {Wang}\ \emph
  {et~al.}(2013{\natexlab{a}})\citenamefont {Wang}, \citenamefont {Lian},
  \citenamefont {Zhang}, \citenamefont {Xu},\ and\ \citenamefont
  {Zhang}}]{wang2013a}%
  \BibitemOpen
  \bibfield  {author} {\bibinfo {author} {\bibfnamefont {J.}~\bibnamefont
  {Wang}}, \bibinfo {author} {\bibfnamefont {B.}~\bibnamefont {Lian}}, \bibinfo
  {author} {\bibfnamefont {H.}~\bibnamefont {Zhang}}, \bibinfo {author}
  {\bibfnamefont {Y.}~\bibnamefont {Xu}}, \ and\ \bibinfo {author}
  {\bibfnamefont {S.-C.}\ \bibnamefont {Zhang}},\ }\href {\doibase
  10.1103/PhysRevLett.111.136801} {\bibfield  {journal} {\bibinfo  {journal}
  {Phys. Rev. Lett.}\ }\textbf {\bibinfo {volume} {111}},\ \bibinfo {pages}
  {136801} (\bibinfo {year} {2013}{\natexlab{a}})}\BibitemShut {NoStop}%
\bibitem [{\citenamefont {Wang}\ \emph
  {et~al.}(2013{\natexlab{b}})\citenamefont {Wang}, \citenamefont {Lian},
  \citenamefont {Zhang},\ and\ \citenamefont {Zhang}}]{wang2013b}%
  \BibitemOpen
  \bibfield  {author} {\bibinfo {author} {\bibfnamefont {J.}~\bibnamefont
  {Wang}}, \bibinfo {author} {\bibfnamefont {B.}~\bibnamefont {Lian}}, \bibinfo
  {author} {\bibfnamefont {H.}~\bibnamefont {Zhang}}, \ and\ \bibinfo {author}
  {\bibfnamefont {S.-C.}\ \bibnamefont {Zhang}},\ }\href {\doibase
  10.1103/PhysRevLett.111.086803} {\bibfield  {journal} {\bibinfo  {journal}
  {Phys. Rev. Lett.}\ }\textbf {\bibinfo {volume} {111}},\ \bibinfo {pages}
  {086803} (\bibinfo {year} {2013}{\natexlab{b}})}\BibitemShut {NoStop}%
\bibitem [{\citenamefont {Zhang}\ \emph {et~al.}(2014)\citenamefont {Zhang},
  \citenamefont {Wang}, \citenamefont {Xu}, \citenamefont {Xu},\ and\
  \citenamefont {Zhang}}]{zhang2014}%
  \BibitemOpen
  \bibfield  {author} {\bibinfo {author} {\bibfnamefont {H.}~\bibnamefont
  {Zhang}}, \bibinfo {author} {\bibfnamefont {J.}~\bibnamefont {Wang}},
  \bibinfo {author} {\bibfnamefont {G.}~\bibnamefont {Xu}}, \bibinfo {author}
  {\bibfnamefont {Y.}~\bibnamefont {Xu}}, \ and\ \bibinfo {author}
  {\bibfnamefont {S.-C.}\ \bibnamefont {Zhang}},\ }\href {\doibase
  10.1103/PhysRevLett.112.096804} {\bibfield  {journal} {\bibinfo  {journal}
  {Phys. Rev. Lett.}\ }\textbf {\bibinfo {volume} {112}},\ \bibinfo {pages}
  {096804} (\bibinfo {year} {2014})}\BibitemShut {NoStop}%
\bibitem [{\citenamefont {Garrity}\ and\ \citenamefont
  {Vanderbilt}(2014)}]{Garrity2014}%
  \BibitemOpen
  \bibfield  {author} {\bibinfo {author} {\bibfnamefont {K.~F.}\ \bibnamefont
  {Garrity}}\ and\ \bibinfo {author} {\bibfnamefont {D.}~\bibnamefont
  {Vanderbilt}},\ }\href {\doibase 10.1103/PhysRevB.90.121103} {\bibfield
  {journal} {\bibinfo  {journal} {Phys. Rev. B}\ }\textbf {\bibinfo {volume}
  {90}},\ \bibinfo {pages} {121103} (\bibinfo {year} {2014})}\BibitemShut
  {NoStop}%
\bibitem [{\citenamefont {Wang}\ \emph {et~al.}(2014)\citenamefont {Wang},
  \citenamefont {Lian},\ and\ \citenamefont {Zhang}}]{wang2014}%
  \BibitemOpen
  \bibfield  {author} {\bibinfo {author} {\bibfnamefont {J.}~\bibnamefont
  {Wang}}, \bibinfo {author} {\bibfnamefont {B.}~\bibnamefont {Lian}}, \ and\
  \bibinfo {author} {\bibfnamefont {S.-C.}\ \bibnamefont {Zhang}},\ }\href
  {\doibase 10.1103/PhysRevB.89.085106} {\bibfield  {journal} {\bibinfo
  {journal} {Phys. Rev. B}\ }\textbf {\bibinfo {volume} {89}},\ \bibinfo
  {pages} {085106} (\bibinfo {year} {2014})}\BibitemShut {NoStop}%
\bibitem [{\citenamefont {Kou}\ \emph {et~al.}(2014)\citenamefont {Kou},
  \citenamefont {Guo}, \citenamefont {Fan}, \citenamefont {Pan}, \citenamefont
  {Lang}, \citenamefont {Jiang}, \citenamefont {Shao}, \citenamefont {Nie},
  \citenamefont {Murata}, \citenamefont {Tang}, \citenamefont {Wang},
  \citenamefont {He}, \citenamefont {Lee}, \citenamefont {Lee},\ and\
  \citenamefont {Wang}}]{kou2014}%
  \BibitemOpen
  \bibfield  {author} {\bibinfo {author} {\bibfnamefont {X.}~\bibnamefont
  {Kou}}, \bibinfo {author} {\bibfnamefont {S.-T.}\ \bibnamefont {Guo}},
  \bibinfo {author} {\bibfnamefont {Y.}~\bibnamefont {Fan}}, \bibinfo {author}
  {\bibfnamefont {L.}~\bibnamefont {Pan}}, \bibinfo {author} {\bibfnamefont
  {M.}~\bibnamefont {Lang}}, \bibinfo {author} {\bibfnamefont {Y.}~\bibnamefont
  {Jiang}}, \bibinfo {author} {\bibfnamefont {Q.}~\bibnamefont {Shao}},
  \bibinfo {author} {\bibfnamefont {T.}~\bibnamefont {Nie}}, \bibinfo {author}
  {\bibfnamefont {K.}~\bibnamefont {Murata}}, \bibinfo {author} {\bibfnamefont
  {J.}~\bibnamefont {Tang}}, \bibinfo {author} {\bibfnamefont {Y.}~\bibnamefont
  {Wang}}, \bibinfo {author} {\bibfnamefont {L.}~\bibnamefont {He}}, \bibinfo
  {author} {\bibfnamefont {T.-K.}\ \bibnamefont {Lee}}, \bibinfo {author}
  {\bibfnamefont {W.-L.}\ \bibnamefont {Lee}}, \ and\ \bibinfo {author}
  {\bibfnamefont {K.~L.}\ \bibnamefont {Wang}},\ }\href {\doibase
  10.1103/PhysRevLett.113.137201} {\bibfield  {journal} {\bibinfo  {journal}
  {Phys. Rev. Lett.}\ }\textbf {\bibinfo {volume} {113}},\ \bibinfo {pages}
  {137201} (\bibinfo {year} {2014})}\BibitemShut {NoStop}%
\bibitem [{\citenamefont {Checkelsky}\ \emph {et~al.}(2014)\citenamefont
  {Checkelsky}, \citenamefont {Yoshimi}, \citenamefont {Tsukazaki},
  \citenamefont {Takahashi}, \citenamefont {Kozuka}, \citenamefont {Falson},
  \citenamefont {Kawasaki},\ and\ \citenamefont {Tokura}}]{checkelsky2014}%
  \BibitemOpen
  \bibfield  {author} {\bibinfo {author} {\bibfnamefont {J.~G.}\ \bibnamefont
  {Checkelsky}}, \bibinfo {author} {\bibfnamefont {R.}~\bibnamefont {Yoshimi}},
  \bibinfo {author} {\bibfnamefont {A.}~\bibnamefont {Tsukazaki}}, \bibinfo
  {author} {\bibfnamefont {K.~S.}\ \bibnamefont {Takahashi}}, \bibinfo {author}
  {\bibfnamefont {Y.}~\bibnamefont {Kozuka}}, \bibinfo {author} {\bibfnamefont
  {J.}~\bibnamefont {Falson}}, \bibinfo {author} {\bibfnamefont
  {M.}~\bibnamefont {Kawasaki}}, \ and\ \bibinfo {author} {\bibfnamefont
  {Y.}~\bibnamefont {Tokura}},\ }\href@noop {} {\bibfield  {journal} {\bibinfo
  {journal} {Nat. Phys.}\ }\textbf {\bibinfo {volume} {10}},\ \bibinfo {pages}
  {731} (\bibinfo {year} {2014})}\BibitemShut {NoStop}%
\bibitem [{\citenamefont {Xu}\ \emph {et~al.}(2015)\citenamefont {Xu},
  \citenamefont {Wang}, \citenamefont {Felser}, \citenamefont {Qi},\ and\
  \citenamefont {Zhang}}]{Xu2015}%
  \BibitemOpen
  \bibfield  {author} {\bibinfo {author} {\bibfnamefont {G.}~\bibnamefont
  {Xu}}, \bibinfo {author} {\bibfnamefont {J.}~\bibnamefont {Wang}}, \bibinfo
  {author} {\bibfnamefont {C.}~\bibnamefont {Felser}}, \bibinfo {author}
  {\bibfnamefont {X.-L.}\ \bibnamefont {Qi}}, \ and\ \bibinfo {author}
  {\bibfnamefont {S.-C.}\ \bibnamefont {Zhang}},\ }\href {\doibase
  10.1021/nl504871u} {\bibfield  {journal} {\bibinfo  {journal} {Nano Letters}\
  }\textbf {\bibinfo {volume} {15}},\ \bibinfo {pages} {2019} (\bibinfo {year}
  {2015})}\BibitemShut {NoStop}%
\bibitem [{\citenamefont {Chang}\ \emph {et~al.}(2015)\citenamefont {Chang},
  \citenamefont {Zhao}, \citenamefont {Kim}, \citenamefont {Zhang},
  \citenamefont {Assaf}, \citenamefont {Heiman}, \citenamefont {Zhang},
  \citenamefont {Liu}, \citenamefont {Chan},\ and\ \citenamefont
  {Moodera}}]{Chang2015}%
  \BibitemOpen
  \bibfield  {author} {\bibinfo {author} {\bibfnamefont {C.-Z.}\ \bibnamefont
  {Chang}}, \bibinfo {author} {\bibfnamefont {W.}~\bibnamefont {Zhao}},
  \bibinfo {author} {\bibfnamefont {D.~Y.}\ \bibnamefont {Kim}}, \bibinfo
  {author} {\bibfnamefont {H.}~\bibnamefont {Zhang}}, \bibinfo {author}
  {\bibfnamefont {B.~A.}\ \bibnamefont {Assaf}}, \bibinfo {author}
  {\bibfnamefont {D.}~\bibnamefont {Heiman}}, \bibinfo {author} {\bibfnamefont
  {S.-C.}\ \bibnamefont {Zhang}}, \bibinfo {author} {\bibfnamefont
  {C.}~\bibnamefont {Liu}}, \bibinfo {author} {\bibfnamefont {M.~H.~W.}\
  \bibnamefont {Chan}}, \ and\ \bibinfo {author} {\bibfnamefont {J.~S.}\
  \bibnamefont {Moodera}},\ }\href@noop {} {\bibfield  {journal} {\bibinfo
  {journal} {Nature Mater.}\ }\textbf {\bibinfo {volume} {14}},\ \bibinfo
  {pages} {473} (\bibinfo {year} {2015})}\BibitemShut {NoStop}%
\bibitem [{\citenamefont {Zhang}\ and\ \citenamefont
  {Zhang}(2012)}]{Zhang2012}%
  \BibitemOpen
  \bibfield  {author} {\bibinfo {author} {\bibfnamefont {X.}~\bibnamefont
  {Zhang}}\ and\ \bibinfo {author} {\bibfnamefont {S.-C.}\ \bibnamefont
  {Zhang}},\ }\href {\doibase 10.1117/12.920325} {\bibfield  {journal}
  {\bibinfo  {journal} {Proc. SPIE}\ }\textbf {\bibinfo {volume} {8373}},\
  \bibinfo {pages} {837309} (\bibinfo {year} {2012})}\BibitemShut {NoStop}%
\bibitem [{\citenamefont {Feng}\ \emph {et~al.}(2015)\citenamefont {Feng},
  \citenamefont {Feng}, \citenamefont {Ou}, \citenamefont {Wang}, \citenamefont
  {Liu}, \citenamefont {Zhang}, \citenamefont {Zhao}, \citenamefont {Jiang},
  \citenamefont {Zhang}, \citenamefont {He}, \citenamefont {Ma}, \citenamefont
  {Xue},\ and\ \citenamefont {Wang}}]{Feng2015}%
  \BibitemOpen
  \bibfield  {author} {\bibinfo {author} {\bibfnamefont {Y.}~\bibnamefont
  {Feng}}, \bibinfo {author} {\bibfnamefont {X.}~\bibnamefont {Feng}}, \bibinfo
  {author} {\bibfnamefont {Y.}~\bibnamefont {Ou}}, \bibinfo {author}
  {\bibfnamefont {J.}~\bibnamefont {Wang}}, \bibinfo {author} {\bibfnamefont
  {C.}~\bibnamefont {Liu}}, \bibinfo {author} {\bibfnamefont {L.}~\bibnamefont
  {Zhang}}, \bibinfo {author} {\bibfnamefont {D.}~\bibnamefont {Zhao}},
  \bibinfo {author} {\bibfnamefont {G.}~\bibnamefont {Jiang}}, \bibinfo
  {author} {\bibfnamefont {S.-C.}\ \bibnamefont {Zhang}}, \bibinfo {author}
  {\bibfnamefont {K.}~\bibnamefont {He}}, \bibinfo {author} {\bibfnamefont
  {X.}~\bibnamefont {Ma}}, \bibinfo {author} {\bibfnamefont {Q.-K.}\
  \bibnamefont {Xue}}, \ and\ \bibinfo {author} {\bibfnamefont
  {Y.}~\bibnamefont {Wang}},\ }\href {\doibase 10.1103/PhysRevLett.115.126801} {\bibfield  {journal} {\bibinfo  {journal}
  {Phys. Rev. Lett.}\ }\textbf {\bibinfo {volume} {115}},\ \bibinfo {pages}
  {126801} (\bibinfo {year} {2015})}\BibitemShut {NoStop}%
  \bibitem [{\citenamefont {Qiao}\ \emph {et~al.}(2010)\citenamefont {Qiao},
  \citenamefont {Yang}, \citenamefont {Feng}, \citenamefont {Tse},
  \citenamefont {Ding}, \citenamefont {Yao}, \citenamefont {Wang},\ and\
  \citenamefont {Niu}}]{Qiao2010}%
  \BibitemOpen
  \bibfield  {author} {\bibinfo {author} {\bibfnamefont {Z.}~\bibnamefont
  {Qiao}}, \bibinfo {author} {\bibfnamefont {S.~A.}\ \bibnamefont {Yang}},
  \bibinfo {author} {\bibfnamefont {W.}~\bibnamefont {Feng}}, \bibinfo {author}
  {\bibfnamefont {W.-K.}\ \bibnamefont {Tse}}, \bibinfo {author} {\bibfnamefont
  {J.}~\bibnamefont {Ding}}, \bibinfo {author} {\bibfnamefont {Y.}~\bibnamefont
  {Yao}}, \bibinfo {author} {\bibfnamefont {J.}~\bibnamefont {Wang}}, \ and\
  \bibinfo {author} {\bibfnamefont {Q.}~\bibnamefont {Niu}},\ }\href {\doibase
  10.1103/PhysRevB.82.161414} {\bibfield  {journal} {\bibinfo  {journal} {Phys.
  Rev. B}\ }\textbf {\bibinfo {volume} {82}},\ \bibinfo {pages} {161414}
  (\bibinfo {year} {2010})}\BibitemShut {NoStop}%
\bibitem [{\citenamefont {Zhang}\ \emph {et~al.}(2012)\citenamefont {Zhang},
  \citenamefont {Lazo}, \citenamefont {Bl\"ugel}, \citenamefont {Heinze},\ and\
  \citenamefont {Mokrousov}}]{Zhang2012b}%
  \BibitemOpen
  \bibfield  {author} {\bibinfo {author} {\bibfnamefont {H.}~\bibnamefont
  {Zhang}}, \bibinfo {author} {\bibfnamefont {C.}~\bibnamefont {Lazo}},
  \bibinfo {author} {\bibfnamefont {S.}~\bibnamefont {Bl\"ugel}}, \bibinfo
  {author} {\bibfnamefont {S.}~\bibnamefont {Heinze}}, \ and\ \bibinfo {author}
  {\bibfnamefont {Y.}~\bibnamefont {Mokrousov}},\ }\href {\doibase
  10.1103/PhysRevLett.108.056802} {\bibfield  {journal} {\bibinfo  {journal}
  {Phys. Rev. Lett.}\ }\textbf {\bibinfo {volume} {108}},\ \bibinfo {pages}
  {056802} (\bibinfo {year} {2012})}\BibitemShut {NoStop}%
\bibitem [{\citenamefont {Ezawa}(2012)}]{Ezawa2012}%
  \BibitemOpen
  \bibfield  {author} {\bibinfo {author} {\bibfnamefont {M.}~\bibnamefont
  {Ezawa}},\ }\href {\doibase 10.1103/PhysRevLett.109.055502} {\bibfield
  {journal} {\bibinfo  {journal} {Phys. Rev. Lett.}\ }\textbf {\bibinfo
  {volume} {109}},\ \bibinfo {pages} {055502} (\bibinfo {year}
  {2012})}\BibitemShut {NoStop}%
\bibitem [{\citenamefont {Qiao}\ \emph {et~al.}(2014)\citenamefont {Qiao},
  \citenamefont {Ren}, \citenamefont {Chen}, \citenamefont {Bellaiche},
  \citenamefont {Zhang}, \citenamefont {MacDonald},\ and\ \citenamefont
  {Niu}}]{Qiao2014}%
  \BibitemOpen
  \bibfield  {author} {\bibinfo {author} {\bibfnamefont {Z.}~\bibnamefont
  {Qiao}}, \bibinfo {author} {\bibfnamefont {W.}~\bibnamefont {Ren}}, \bibinfo
  {author} {\bibfnamefont {H.}~\bibnamefont {Chen}}, \bibinfo {author}
  {\bibfnamefont {L.}~\bibnamefont {Bellaiche}}, \bibinfo {author}
  {\bibfnamefont {Z.}~\bibnamefont {Zhang}}, \bibinfo {author} {\bibfnamefont
  {A.~H.}\ \bibnamefont {MacDonald}}, \ and\ \bibinfo {author} {\bibfnamefont
  {Q.}~\bibnamefont {Niu}},\ }\href {\doibase 10.1103/PhysRevLett.112.116404}
  {\bibfield  {journal} {\bibinfo  {journal} {Phys. Rev. Lett.}\ }\textbf
  {\bibinfo {volume} {112}},\ \bibinfo {pages} {116404} (\bibinfo {year}
  {2014})}\BibitemShut {NoStop}%
\bibitem [{\citenamefont {Wang}\ \emph {et~al.}(2015)\citenamefont {Wang},
  \citenamefont {Tang}, \citenamefont {Sachs}, \citenamefont {Barlas},\ and\
  \citenamefont {Shi}}]{Wang2015}%
  \BibitemOpen
  \bibfield  {author} {\bibinfo {author} {\bibfnamefont {Z.}~\bibnamefont
  {Wang}}, \bibinfo {author} {\bibfnamefont {C.}~\bibnamefont {Tang}}, \bibinfo
  {author} {\bibfnamefont {R.}~\bibnamefont {Sachs}}, \bibinfo {author}
  {\bibfnamefont {Y.}~\bibnamefont {Barlas}}, \ and\ \bibinfo {author}
  {\bibfnamefont {J.}~\bibnamefont {Shi}},\ }\href {\doibase
  10.1103/PhysRevLett.114.016603} {\bibfield  {journal} {\bibinfo  {journal}
  {Phys. Rev. Lett.}\ }\textbf {\bibinfo {volume} {114}},\ \bibinfo {pages}
  {016603} (\bibinfo {year} {2015})}\BibitemShut {NoStop}%
\bibitem [{\citenamefont {Ohgushi}\ \emph {et~al.}(2000)\citenamefont
  {Ohgushi}, \citenamefont {Murakami},\ and\ \citenamefont
  {Nagaosa}}]{Ohgushi2000}%
  \BibitemOpen
  \bibfield  {author} {\bibinfo {author} {\bibfnamefont {K.}~\bibnamefont
  {Ohgushi}}, \bibinfo {author} {\bibfnamefont {S.}~\bibnamefont {Murakami}}, \
  and\ \bibinfo {author} {\bibfnamefont {N.}~\bibnamefont {Nagaosa}},\ }\href
  {\doibase 10.1103/PhysRevB.62.R6065} {\bibfield  {journal} {\bibinfo
  {journal} {Phys. Rev. B}\ }\textbf {\bibinfo {volume} {62}},\ \bibinfo
  {pages} {R6065} (\bibinfo {year} {2000})}\BibitemShut {NoStop}%
\bibitem [{\citenamefont {Zhang}(2011)}]{Zhang2011}%
  \BibitemOpen
  \bibfield  {author} {\bibinfo {author} {\bibfnamefont {Z.-Y.}\ \bibnamefont
  {Zhang}},\ }\href {http://stacks.iop.org/0953-8984/23/i=36/a=365801}
  {\bibfield  {journal} {\bibinfo  {journal} {Journal of Physics: Condensed
  Matter}\ }\textbf {\bibinfo {volume} {23}},\ \bibinfo {pages} {365801}
  (\bibinfo {year} {2011})}\BibitemShut {NoStop}%
\bibitem [{\citenamefont {Englich}\ \emph {et~al.}(1997)\citenamefont
  {Englich}, \citenamefont {Frommen},\ and\ \citenamefont
  {Massa}}]{Englich1997}%
  \BibitemOpen
  \bibfield  {author} {\bibinfo {author} {\bibfnamefont {U.}~\bibnamefont
  {Englich}}, \bibinfo {author} {\bibfnamefont {C.}~\bibnamefont {Frommen}}, \
  and\ \bibinfo {author} {\bibfnamefont {W.}~\bibnamefont {Massa}},\ }\href
  {\doibase http://dx.doi.org/10.1016/S0925-8388(96)02478-4} {\bibfield
  {journal} {\bibinfo  {journal} {Journal of Alloys and Compounds}\ }\textbf
  {\bibinfo {volume} {246}},\ \bibinfo {pages} {155 } (\bibinfo {year}
  {1997})}\BibitemShut {NoStop}%
\bibitem [{\citenamefont {Shores}\ \emph {et~al.}(2005)\citenamefont {Shores},
  \citenamefont {Nytko}, \citenamefont {Bartlett},\ and\ \citenamefont
  {Nocera}}]{Shores2005}%
  \BibitemOpen
  \bibfield  {author} {\bibinfo {author} {\bibfnamefont {M.~P.}\ \bibnamefont
  {Shores}}, \bibinfo {author} {\bibfnamefont {E.~A.}\ \bibnamefont {Nytko}},
  \bibinfo {author} {\bibfnamefont {B.~M.}\ \bibnamefont {Bartlett}}, \ and\
  \bibinfo {author} {\bibfnamefont {D.~G.}\ \bibnamefont {Nocera}},\ }\href
  {\doibase 10.1021/ja053891p} {\bibfield  {journal} {\bibinfo  {journal}
  {Journal of the American Chemical Society}\ }\textbf {\bibinfo {volume}
  {127}},\ \bibinfo {pages} {13462} (\bibinfo {year} {2005})}\BibitemShut
  {NoStop}%
\bibitem [{\citenamefont {Wang}\ \emph
  {et~al.}(2013{\natexlab{c}})\citenamefont {Wang}, \citenamefont {Su},\ and\
  \citenamefont {Liu}}]{Wang2013c}%
  \BibitemOpen
  \bibfield  {author} {\bibinfo {author} {\bibfnamefont {Z.~F.}\ \bibnamefont
  {Wang}}, \bibinfo {author} {\bibfnamefont {N.}~\bibnamefont {Su}}, \ and\
  \bibinfo {author} {\bibfnamefont {F.}~\bibnamefont {Liu}},\ }\href {\doibase
  10.1021/nl401147u} {\bibfield  {journal} {\bibinfo  {journal} {Nano Letters}\
  }\textbf {\bibinfo {volume} {13}},\ \bibinfo {pages} {2842} (\bibinfo {year}
  {2013}{\natexlab{c}})}\BibitemShut {NoStop}%
\bibitem [{\citenamefont {Tang}\ \emph {et~al.}(2011)\citenamefont {Tang},
  \citenamefont {Mei},\ and\ \citenamefont {Wen}}]{Tang2011}%
  \BibitemOpen
  \bibfield  {author} {\bibinfo {author} {\bibfnamefont {E.}~\bibnamefont
  {Tang}}, \bibinfo {author} {\bibfnamefont {J.-W.}\ \bibnamefont {Mei}}, \
  and\ \bibinfo {author} {\bibfnamefont {X.-G.}\ \bibnamefont {Wen}},\ }\href
  {\doibase 10.1103/PhysRevLett.106.236802} {\bibfield  {journal} {\bibinfo
  {journal} {Phys. Rev. Lett.}\ }\textbf {\bibinfo {volume} {106}},\ \bibinfo
  {pages} {236802} (\bibinfo {year} {2011})}\BibitemShut {NoStop}%
\bibitem [{\citenamefont {Venderbos}\ \emph {et~al.}(2011)\citenamefont
  {Venderbos}, \citenamefont {Daghofer},\ and\ \citenamefont {van~den
  Brink}}]{Venderbos2011}%
  \BibitemOpen
  \bibfield  {author} {\bibinfo {author} {\bibfnamefont {J.~W.~F.}\
  \bibnamefont {Venderbos}}, \bibinfo {author} {\bibfnamefont {M.}~\bibnamefont
  {Daghofer}}, \ and\ \bibinfo {author} {\bibfnamefont {J.}~\bibnamefont
  {van~den Brink}},\ }\href {\doibase 10.1103/PhysRevLett.107.116401}
  {\bibfield  {journal} {\bibinfo  {journal} {Phys. Rev. Lett.}\ }\textbf
  {\bibinfo {volume} {107}},\ \bibinfo {pages} {116401} (\bibinfo {year}
  {2011})}\BibitemShut {NoStop}%
\bibitem [{\citenamefont {Nishimoto}\ \emph {et~al.}(2010)\citenamefont
  {Nishimoto}, \citenamefont {Nakamura}, \citenamefont {O'Brien},\ and\
  \citenamefont {Fulde}}]{Nishimoto2010}%
  \BibitemOpen
  \bibfield  {author} {\bibinfo {author} {\bibfnamefont {S.}~\bibnamefont
  {Nishimoto}}, \bibinfo {author} {\bibfnamefont {M.}~\bibnamefont {Nakamura}},
  \bibinfo {author} {\bibfnamefont {A.}~\bibnamefont {O'Brien}}, \ and\
  \bibinfo {author} {\bibfnamefont {P.}~\bibnamefont {Fulde}},\ }\href
  {\doibase 10.1103/PhysRevLett.104.196401} {\bibfield  {journal} {\bibinfo
  {journal} {Phys. Rev. Lett.}\ }\textbf {\bibinfo {volume} {104}},\ \bibinfo
  {pages} {196401} (\bibinfo {year} {2010})}\BibitemShut {NoStop}%
\bibitem [{\citenamefont {Green}\ \emph {et~al.}(2010)\citenamefont {Green},
  \citenamefont {Santos},\ and\ \citenamefont {Chamon}}]{Green2010}%
  \BibitemOpen
  \bibfield  {author} {\bibinfo {author} {\bibfnamefont {D.}~\bibnamefont
  {Green}}, \bibinfo {author} {\bibfnamefont {L.}~\bibnamefont {Santos}}, \
  and\ \bibinfo {author} {\bibfnamefont {C.}~\bibnamefont {Chamon}},\ }\href
  {\doibase 10.1103/PhysRevB.82.075104} {\bibfield  {journal} {\bibinfo
  {journal} {Phys. Rev. B}\ }\textbf {\bibinfo {volume} {82}},\ \bibinfo
  {pages} {075104} (\bibinfo {year} {2010})}\BibitemShut {NoStop}%
\bibitem [{\citenamefont {Liu}\ \emph {et~al.}(2010)\citenamefont {Liu},
  \citenamefont {Yao},\ and\ \citenamefont {Ma}}]{Liu2010b}%
  \BibitemOpen
  \bibfield  {author} {\bibinfo {author} {\bibfnamefont {Q.}~\bibnamefont
  {Liu}}, \bibinfo {author} {\bibfnamefont {H.}~\bibnamefont {Yao}}, \ and\
  \bibinfo {author} {\bibfnamefont {T.}~\bibnamefont {Ma}},\ }\href {\doibase
  10.1103/PhysRevB.82.045102} {\bibfield  {journal} {\bibinfo  {journal} {Phys.
  Rev. B}\ }\textbf {\bibinfo {volume} {82}},\ \bibinfo {pages} {045102}
  (\bibinfo {year} {2010})}\BibitemShut {NoStop}%
\bibitem [{\citenamefont {Nakata}\ \emph {et~al.}(2012)\citenamefont {Nakata},
  \citenamefont {Okada}, \citenamefont {Nakanishi},\ and\ \citenamefont
  {Kitano}}]{Nakata2012}%
  \BibitemOpen
  \bibfield  {author} {\bibinfo {author} {\bibfnamefont {Y.}~\bibnamefont
  {Nakata}}, \bibinfo {author} {\bibfnamefont {T.}~\bibnamefont {Okada}},
  \bibinfo {author} {\bibfnamefont {T.}~\bibnamefont {Nakanishi}}, \ and\
  \bibinfo {author} {\bibfnamefont {M.}~\bibnamefont {Kitano}},\ }\href
  {\doibase 10.1103/PhysRevB.85.205128} {\bibfield  {journal} {\bibinfo
  {journal} {Phys. Rev. B}\ }\textbf {\bibinfo {volume} {85}},\ \bibinfo
  {pages} {205128} (\bibinfo {year} {2012})}\BibitemShut {NoStop}%
\bibitem [{\citenamefont {Massa}\ and\ \citenamefont
  {Steiner}(1980)}]{Massa1980}%
  \BibitemOpen
  \bibfield  {author} {\bibinfo {author} {\bibfnamefont {W.}~\bibnamefont
  {Massa}}\ and\ \bibinfo {author} {\bibfnamefont {M.}~\bibnamefont
  {Steiner}},\ }\href {\doibase http://dx.doi.org/10.1016/0022-4596(80)90559-9}
  {\bibfield  {journal} {\bibinfo  {journal} {Journal of Solid State
  Chemistry}\ }\textbf {\bibinfo {volume} {32}},\ \bibinfo {pages} {137 }
  (\bibinfo {year} {1980})}\BibitemShut {NoStop}%
\bibitem [{\citenamefont {Hohenberg}\ and\ \citenamefont
  {Kohn}(1964)}]{Hohenberg1964}%
  \BibitemOpen
  \bibfield  {author} {\bibinfo {author} {\bibfnamefont {P.}~\bibnamefont
  {Hohenberg}}\ and\ \bibinfo {author} {\bibfnamefont {W.}~\bibnamefont
  {Kohn}},\ }\href {\doibase 10.1103/PhysRev.136.B864} {\bibfield  {journal}
  {\bibinfo  {journal} {Phys. Rev.}\ }\textbf {\bibinfo {volume} {136}},\
  \bibinfo {pages} {B864} (\bibinfo {year} {1964})}\BibitemShut {NoStop}%
\bibitem [{\citenamefont {Kohn}\ and\ \citenamefont {Sham}(1965)}]{Kohn1965}%
  \BibitemOpen
  \bibfield  {author} {\bibinfo {author} {\bibfnamefont {W.}~\bibnamefont
  {Kohn}}\ and\ \bibinfo {author} {\bibfnamefont {L.~J.}\ \bibnamefont
  {Sham}},\ }\href {\doibase 10.1103/PhysRev.140.A1133} {\bibfield  {journal}
  {\bibinfo  {journal} {Phys. Rev.}\ }\textbf {\bibinfo {volume} {140}},\
  \bibinfo {pages} {A1133} (\bibinfo {year} {1965})}\BibitemShut {NoStop}%
\bibitem [{\citenamefont {Vanderbilt}(1990)}]{Vanderbilt1990}%
  \BibitemOpen
  \bibfield  {author} {\bibinfo {author} {\bibfnamefont {D.}~\bibnamefont
  {Vanderbilt}},\ }\href {\doibase 10.1103/PhysRevB.41.7892} {\bibfield
  {journal} {\bibinfo  {journal} {Phys. Rev. B}\ }\textbf {\bibinfo {volume}
  {41}},\ \bibinfo {pages} {7892} (\bibinfo {year} {1990})}\BibitemShut
  {NoStop}%
\bibitem [{\citenamefont {Perdew}\ \emph {et~al.}(1996)\citenamefont {Perdew},
  \citenamefont {Burke},\ and\ \citenamefont {Ernzerhof}}]{Perdew1996}%
  \BibitemOpen
  \bibfield  {author} {\bibinfo {author} {\bibfnamefont {J.~P.}\ \bibnamefont
  {Perdew}}, \bibinfo {author} {\bibfnamefont {K.}~\bibnamefont {Burke}}, \
  and\ \bibinfo {author} {\bibfnamefont {M.}~\bibnamefont {Ernzerhof}},\ }\href
  {\doibase 10.1103/PhysRevLett.77.3865} {\bibfield  {journal} {\bibinfo
  {journal} {Phys. Rev. Lett.}\ }\textbf {\bibinfo {volume} {77}},\ \bibinfo
  {pages} {3865} (\bibinfo {year} {1996})}\BibitemShut {NoStop}%
\bibitem [{\citenamefont {Fang}\ and\ \citenamefont
  {Terakura}(2002)}]{Fang2002}%
  \BibitemOpen
  \bibfield  {author} {\bibinfo {author} {\bibfnamefont {Z.}~\bibnamefont
  {Fang}}\ and\ \bibinfo {author} {\bibfnamefont {K.}~\bibnamefont
  {Terakura}},\ }\href {http://stacks.iop.org/0953-8984/14/i=11/a=312}
  {\bibfield  {journal} {\bibinfo  {journal} {Journal of Physics: Condensed
  Matter}\ }\textbf {\bibinfo {volume} {14}},\ \bibinfo {pages} {3001}
  (\bibinfo {year} {2002})}\BibitemShut {NoStop}%
\bibitem [{\citenamefont {Nordheim}(1931)}]{Nordheim1931}%
  \BibitemOpen
  \bibfield  {author} {\bibinfo {author} {\bibfnamefont {L.}~\bibnamefont
  {Nordheim}},\ }\href {\doibase 10.1002/andp.19314010507} {\bibfield
  {journal} {\bibinfo  {journal} {Ann. Phys}\ }\textbf {\bibinfo
  {volume} {9}},\ \bibinfo {pages} {607} (\bibinfo {year}
  {1931})}\BibitemShut {NoStop}%
\bibitem [{\citenamefont {Bellaiche}\ and\ \citenamefont
  {Vanderbilt}(2000)}]{Bellaiche2000}%
  \BibitemOpen
  \bibfield  {author} {\bibinfo {author} {\bibfnamefont {L.}~\bibnamefont
  {Bellaiche}}\ and\ \bibinfo {author} {\bibfnamefont {D.}~\bibnamefont
  {Vanderbilt}},\ }\href {\doibase 10.1103/PhysRevB.61.7877} {\bibfield
  {journal} {\bibinfo  {journal} {Phys. Rev. B}\ }\textbf {\bibinfo {volume}
  {61}},\ \bibinfo {pages} {7877} (\bibinfo {year} {2000})}\BibitemShut
  {NoStop}%
\bibitem [{\citenamefont {Kresse}\ and\ \citenamefont
  {Hafner}(1993)}]{Kresse1993}%
  \BibitemOpen
  \bibfield  {author} {\bibinfo {author} {\bibfnamefont {G.}~\bibnamefont
  {Kresse}}\ and\ \bibinfo {author} {\bibfnamefont {J.}~\bibnamefont
  {Hafner}},\ }\href {\doibase 10.1103/PhysRevB.47.558} {\bibfield  {journal}
  {\bibinfo  {journal} {Phys. Rev. B}\ }\textbf {\bibinfo {volume} {47}},\
  \bibinfo {pages} {558} (\bibinfo {year} {1993})}\BibitemShut {NoStop}%
\bibitem [{\citenamefont {Kresse}\ and\ \citenamefont
  {Furthm\"uller}(1996)}]{Kresse1996}%
  \BibitemOpen
  \bibfield  {author} {\bibinfo {author} {\bibfnamefont {G.}~\bibnamefont
  {Kresse}}\ and\ \bibinfo {author} {\bibfnamefont {J.}~\bibnamefont
  {Furthm\"uller}},\ }\href {\doibase 10.1103/PhysRevB.54.11169} {\bibfield
  {journal} {\bibinfo  {journal} {Phys. Rev. B}\ }\textbf {\bibinfo {volume}
  {54}},\ \bibinfo {pages} {11169} (\bibinfo {year} {1996})}\BibitemShut
  {NoStop}%
\bibitem [{\citenamefont {Jeschke}\ \emph {et~al.}(2013)\citenamefont
  {Jeschke}, \citenamefont {Salvat-Pujol},\ and\ \citenamefont
  {Valent\'{i}}}]{Jeschke2013}%
  \BibitemOpen
  \bibfield  {author} {\bibinfo {author} {\bibfnamefont {H.~O.}\ \bibnamefont
  {Jeschke}}, \bibinfo {author} {\bibfnamefont {F.}~\bibnamefont
  {Salvat-Pujol}}, \ and\ \bibinfo {author} {\bibfnamefont {R.}~\bibnamefont
  {Valent\'{i}}},\ }\href {\doibase 10.1103/PhysRevB.88.075106} {\bibfield
  {journal} {\bibinfo  {journal} {Phys. Rev. B}\ }\textbf {\bibinfo {volume}
  {88}},\ \bibinfo {pages} {075106} (\bibinfo {year} {2013})}\BibitemShut
  {NoStop}%
\bibitem [{\citenamefont {Weng}\ \emph {et~al.}(2011)\citenamefont {Weng},
  \citenamefont {Xu}, \citenamefont {Zhang}, \citenamefont {Zhang},
  \citenamefont {Dai},\ and\ \citenamefont {Fang}}]{Weng2011}%
  \BibitemOpen
  \bibfield  {author} {\bibinfo {author} {\bibfnamefont {H.}~\bibnamefont
  {Weng}}, \bibinfo {author} {\bibfnamefont {G.}~\bibnamefont {Xu}}, \bibinfo
  {author} {\bibfnamefont {H.}~\bibnamefont {Zhang}}, \bibinfo {author}
  {\bibfnamefont {S.-C.}\ \bibnamefont {Zhang}}, \bibinfo {author}
  {\bibfnamefont {X.}~\bibnamefont {Dai}}, \ and\ \bibinfo {author}
  {\bibfnamefont {Z.}~\bibnamefont {Fang}},\ }\href {\doibase
  10.1103/PhysRevB.84.060408} {\bibfield  {journal} {\bibinfo  {journal} {Phys.
  Rev. B}\ }\textbf {\bibinfo {volume} {84}},\ \bibinfo {pages} {060408}
  (\bibinfo {year} {2011})}\BibitemShut {NoStop}%
\bibitem [{\citenamefont {Yao}\ \emph {et~al.}(2015)\citenamefont {Yao},
  \citenamefont {Miao}, \citenamefont {Wang}, \citenamefont {Dil},
  \citenamefont {Hasan}, \citenamefont {Guan}, \citenamefont {Gao},
  \citenamefont {Liu}, \citenamefont {Qian},\ and\ \citenamefont
  {Jia}}]{Yao2015}%
  \BibitemOpen
  \bibfield  {author} {\bibinfo {author} {\bibfnamefont {M.-Y.}\ \bibnamefont
  {Yao}}, \bibinfo {author} {\bibfnamefont {L.}~\bibnamefont {Miao}}, \bibinfo
  {author} {\bibfnamefont {N.~L.}\ \bibnamefont {Wang}}, \bibinfo {author}
  {\bibfnamefont {J.~H.}\ \bibnamefont {Dil}}, \bibinfo {author} {\bibfnamefont
  {M.~Z.}\ \bibnamefont {Hasan}}, \bibinfo {author} {\bibfnamefont {D.~D.}\
  \bibnamefont {Guan}}, \bibinfo {author} {\bibfnamefont {C.~L.}\ \bibnamefont
  {Gao}}, \bibinfo {author} {\bibfnamefont {C.}~\bibnamefont {Liu}}, \bibinfo
  {author} {\bibfnamefont {D.}~\bibnamefont {Qian}}, \ and\ \bibinfo {author}
  {\bibfnamefont {J.-f.}\ \bibnamefont {Jia}},\ }\href {\doibase
  10.1103/PhysRevB.91.161411} {\bibfield  {journal} {\bibinfo  {journal} {Phys.
  Rev. B}\ }\textbf {\bibinfo {volume} {91}},\ \bibinfo {pages} {161411}
  (\bibinfo {year} {2015})}\BibitemShut {NoStop}%
\bibitem [{\citenamefont {Kane}\ and\ \citenamefont {Mele}(2005)}]{kane2005a}%
  \BibitemOpen
  \bibfield  {author} {\bibinfo {author} {\bibfnamefont {C.~L.}\ \bibnamefont
  {Kane}}\ and\ \bibinfo {author} {\bibfnamefont {E.~J.}\ \bibnamefont
  {Mele}},\ }\href@noop {} {\bibfield  {journal} {\bibinfo  {journal} {Phys.
  Rev. Lett.}\ }\textbf {\bibinfo {volume} {95}},\ \bibinfo {pages} {226801}
  (\bibinfo {year} {2005})}\BibitemShut {NoStop}%
\bibitem [{\citenamefont {Sancho}\ \emph {et~al.}(1985)\citenamefont {Sancho},
  \citenamefont {Sancho}, \citenamefont {Sancho},\ and\ \citenamefont
  {Rubio}}]{Sancho1985}%
  \BibitemOpen
  \bibfield  {author} {\bibinfo {author} {\bibfnamefont {M.~L.}\ \bibnamefont
  {Sancho}}, \bibinfo {author} {\bibfnamefont {J.~L.}\ \bibnamefont {Sancho}},
  \bibinfo {author} {\bibfnamefont {J.~L.}\ \bibnamefont {Sancho}}, \ and\
  \bibinfo {author} {\bibfnamefont {J.}~\bibnamefont {Rubio}},\ }\href@noop {}
  {\bibfield  {journal} {\bibinfo  {journal} {Journal of Physics F: Metal
  Physics}\ }\textbf {\bibinfo {volume} {15}},\ \bibinfo {pages} {851}
  (\bibinfo {year} {1985})}\BibitemShut {NoStop}%
\bibitem [{\citenamefont {Marzari}\ and\ \citenamefont
  {Vanderbilt}(1997)}]{Marzari1997}%
  \BibitemOpen
  \bibfield  {author} {\bibinfo {author} {\bibfnamefont {N.}~\bibnamefont
  {Marzari}}\ and\ \bibinfo {author} {\bibfnamefont {D.}~\bibnamefont
  {Vanderbilt}},\ }\href {\doibase 10.1103/PhysRevB.56.12847} {\bibfield
  {journal} {\bibinfo  {journal} {Phys. Rev. B}\ }\textbf {\bibinfo {volume}
  {56}},\ \bibinfo {pages} {12847} (\bibinfo {year} {1997})}\BibitemShut
  {NoStop}%
\bibitem [{\citenamefont {Souza}\ \emph {et~al.}(2001)\citenamefont {Souza},
  \citenamefont {Marzari},\ and\ \citenamefont {Vanderbilt}}]{Souza2001}%
  \BibitemOpen
  \bibfield  {author} {\bibinfo {author} {\bibfnamefont {I.}~\bibnamefont
  {Souza}}, \bibinfo {author} {\bibfnamefont {N.}~\bibnamefont {Marzari}}, \
  and\ \bibinfo {author} {\bibfnamefont {D.}~\bibnamefont {Vanderbilt}},\
  }\href {\doibase 10.1103/PhysRevB.65.035109} {\bibfield  {journal} {\bibinfo
  {journal} {Phys. Rev. B}\ }\textbf {\bibinfo {volume} {65}},\ \bibinfo
  {pages} {035109} (\bibinfo {year} {2001})}\BibitemShut {NoStop}%
\end{thebibliography}
\end{document}